\documentclass[submission, Phys]{SciPost}

\usepackage{braket}
\usepackage{nicefrac}
\usepackage{cleveref}
\usepackage{amssymb}

\usepackage{calc}
\usepackage{pgf}
\usepackage{tikz}
\tikzset{>=stealth}
\usepackage{pgffor}
\usepackage{pgfplots}
\pgfplotsset{compat=1.8}
\usepackage{pgfplotstable}
\usepgfplotslibrary{groupplots}
\usepgfplotslibrary{fillbetween}
\usepgfplotslibrary{patchplots}
\usetikzlibrary
{
	calc,
	decorations,
	plotmarks,
	patterns,
	positioning,
	petri,
	arrows,
	intersections,
	decorations.markings,
	backgrounds,
	fit,
	matrix,
	graphs,
	shapes.geometric,
	decorations.pathreplacing, 
	decorations.pathmorphing,
	shapes.misc,
	shapes.multipart,
	shapes,
	through,
	tikzmark,
	arrows.meta
}
\newboolean{buildtikzpics}
\setboolean{buildtikzpics}{false}
\newif\ifrebuildtikz
\newif\ifChangeMode
\ChangeModetrue
\ChangeModefalse
\ifthenelse{\boolean{buildtikzpics}}
{
	\rebuildtikztrue
	\usetikzlibrary{external}
	\tikzexternalize[optimize=false,prefix=figures/autogen/]
}
{
	\rebuildtikzfalse
        \newcommand{\tikzsetnextfilename}[1]{}
}

\crefname{figure}{Fig.}{Figs.}
\Crefname{figure}{Fig.}{Figs.}
\crefname{section}{Sec.}{Secs.}
\Crefname{section}{Sec.}{Secs.}
\begin{document}
\newcommand{\todo}[1]{\textbf{\color{red}{[#1]}}}

\begin{center}{\Large \textbf{ Evaluation of time-dependent
      correlators after a local quench in iPEPS: hole motion in
      the $t-J$ model }}\end{center}

\begin{center}
C. Hubig\textsuperscript{1,4*},
A. Bohrdt\textsuperscript{2,4},
M. Knap\textsuperscript{2,4},
F. Grusdt\textsuperscript{3,4},
J. I. Cirac\textsuperscript{1,4}
\end{center}

\begin{center}
{\bf 1} Max-Planck-Institut für Quantenoptik, 85748 Garching, Germany
\\
{\bf 2} Department of Physics and Institute for Advanced Study,\\Technical University of Munich, 85748 Garching, Germany
\\
{\bf 3} Department of Physics and Arnold Sommerfeld Center for Theoretical Physics (ASC), Ludwig-Maximilians-Universität München, 80333 München, Germany
\\
{\bf 4} Munich Center for Quantum Science and Technology (MCQST),\\80799 München, Germany
\\
* claudius.hubig@mpq.mpg.de
\end{center}

\begin{center}
\today
\end{center}

\section*{Abstract} {\bf Infinite projected entangled pair states
  (iPEPS) provide a convenient variational description of infinite,
  translationally-invariant two-dimensional quantum states. However,
  the simulation of local excitations is not directly possible due to
  the translationally-invariant ansatz. Furthermore, as iPEPS are
  either identical or orthogonal, expectation values between different
  states as required during the evaluation of non-equal-time
  correlators are ill-defined. \\
  Here, we show that by introducing auxiliary states on each site, it
  becomes possible to simulate both local excitations and evaluate
  non-equal-time correlators in an iPEPS setting under real-time
  evolution. We showcase the method by simulating the $t-J$ model
  after a single hole has been placed in the half-filled
  antiferromagnetic background and evaluating both return
  probabilities and spin correlation functions, as accessible in
  quantum gas microscopes.}

\vspace{10pt}
\noindent\rule{\textwidth}{1pt}
\tableofcontents\thispagestyle{fancy}
\noindent\rule{\textwidth}{1pt}
\vspace{10pt}

\section{\label{sec:intro}Introduction}

While tensor network methods in the form of matrix-product states have
become the method of choice for the simulation of one-dimensional
quantum systems and provide both excellent ground-state
data\cite{schollwoeck11} and good accuracy for time-dependent
quantities\cite{paeckel19:_time}, the study of two-dimensional systems
remains more difficult. The limited system size of methods such as
exact diagonalisation or matrix-product states on a
cylinder\cite{stoudenmire12:_study_two_dimen_system_densit} becomes
particularly relevant when studying time-dependent correlators after
local excitations, as the system must be able to accommodate the
spread of those correlations over time and avoid their interaction
with any boundaries. Infinite projected entangled pair
states\cite{verstraete04:_renor, verstraete08:_matrix,
  jordan08:_class_simul_infin_size_quant} (iPEPS) on the other hand
allow for the simulation of ground-state properties of \emph{infinite}
two-dimensional systems with high accuracy by repeating a finite unit
cell of tensors infinitely in both directions.  iPEPS were also
recently shown to allow for the simulation of global
quenches\cite{czarnik19:_time, hubig19:_time, kshetrimayum19:_time} at
least for short times. This simulation of a real-time evolution
following a global quantum quench is relatively straightforward:
evolution methods exist\cite{lubasch14:_algor, phien15:_fast,
  phien15:_infin}, the quench can be enacted by a change of the
Hamiltonian governing this evolution and translational invariance is
retained. Equal-time correlators can also be
evaluated as usual for each of the computed time-evolved post-quench
states.

However, when attempting to simulate a local quench and evaluate
non-equal-time correlators, one encounters two problems: First, it is
not possible to simply apply an operator (such as $\hat c^\dagger$) to
a single site of the quantum state to create the local excitation: To
follow this route, one would have to apply this operator to a specific
site, repeated on each unit cell. While making the unit cell itself
relatively large is feasible, in this case one merely recovers the
case of a finite PEPS calculation and loses the inherent infinity of
the iPEPS ansatz. The handling of fermionic commutation rules further
complicates this approach.

Second, when pursuing this avenue to simulate the evolution of many
excitations -- one per unit cell -- over time, it is then still not
possible to evaluate non-equal-time correlators: These correlators are
calculated as expectation values between two different quantum
states. However, evaluating the norms of those states will yield
either 0 or 1 in the thermodynamic limit and the scale of the
correlator is hence not known. In comparison, equal-time correlators
are evaluated as
$\braket{\hat O(t)} = \frac{\bra{ \psi(t)} \hat O
  \ket{\psi(t)}}{\braket{\psi(t) | \psi(t)}}$,
but the denumerator is clearly ill-defined for a correlator
$\braket{\hat O(t^\prime, t)}$ between two different infinite quantum
states $\Ket{\psi(t^\prime)}$ and $\Ket{\psi(t)}$.

Here, we avoid both problems by adding one auxiliary site to each of
the physical sites of our system while preserving translational
invariance. We demonstrate the method by evaluating the return
probability and diagonal-spin-correlators of a single hole in the
two-dimensional antiferromagnetic background of the $t-J$ model\cite{dagotto90:_stron, poilblanc92:_singl, poilblanc93:_singl, poilblanc93:_dynam, beran96:_eviden_jbb, bohrdt19:_dynam, zhang91:_exact_j_hubbar_hamil, mierzejewski11:_noneq_quant_dynam_charg_carrier, lenarifmmode14:_optic, goleifmmode14:_mechan, eckstein14:_ultraf_separ_photod_carrier_mott_antif}.

\section{\label{sec:exc}Local excitations and non-equal-time
  correlators}

Consider a system composed of physical local state spaces
$\mathcal{H}^p_i$ repeated on each site $i$ of an infinite lattice. We
will later focus on the case of a square two-dimensional lattice, but
the method likewise applies to other lattice geometries.  The total
Hilbert space is the tensor product of the local spaces,
\begin{equation}
  \mathcal{H}^p = \bigotimes_i \mathcal{H}^p_i \;.
\end{equation}
We can represent a translationally invariant quantum state
$\Ket{\psi^p} \in \mathcal{H}^p$ using a tensor network ansatz if it
has low entanglement, which is typically true for ground states of
local Hamiltonians. If $\Ket{\psi^p}$ is only invariant under
translation by multiple sites (such as e.g. an antiferromagnetic state
under translation by two instead of one site), we can also capture
this by using a sufficiently large unit cell of tensors in the ansatz.

To simulate a local excitation without breaking translational
invariance, we now create a translationally invariant superposition of
excitations on top of our initial state, simulate the time evolution
of this superposition under some Hamiltonian $\hat H$ and then select
the part of the superposition which contains an excitation at a
specific local site\cite{paredes05:_exploit_quant_paral_simul_quant,
  knap13:_probin_real_space_time_resol}.

To create the superposition of local excitations, one could apply
e.g. $\left( \hat 1 + \epsilon \hat x^p_i \right)$ with some creation or
annihilation operator $\hat x^p_i$ and a small prefactor $\epsilon$
governing the density of excitations on each site as
\begin{equation}
  \hat Y = \prod_i \left( \hat 1 + \epsilon \hat x^p_i \right) \;.
\end{equation}
If we let this operator act on our quantum state, we obtain a superposition
\begin{equation}
  \hat Y \Ket{\psi^p} = \Ket{\psi^p} + \sum_i \epsilon \hat x^p_i \Ket{\psi^p} + \mathcal{O}(\epsilon^2).
\end{equation}
By including a suitable operator (e.g. the particle number operator)
in expectation values later, we can select one of the states with an
excitation (e.g. a hole at a particular site), which is most likely
one of the summands in the second term if $\epsilon$ is
small. Crucially, we can also do so after a real-time evolution of
$\hat Y \Ket{\psi^p}$, in this way post-selecting the evolution of a
single excitation out of the translationally invariant background.

This approach using $\hat Y$ has two downsides: First, the operator
$\hat x^p_i$ alone typically breaks some symmetry of the system such
as spin projection, particle conservation or fermionic parity. While
the former two merely lead to a less efficient simulation (as those
symmetries then cannot be used in the tensor network ansatz), the
breaking of fermionic parity is a serious problem which makes the
simulation of fermionic systems impossible. Furthermore, while it is
possible to post-select a quantum state with an excitation present at
a particular site \emph{after} the time evolution, we cannot
post-select for a state where the excitation was \emph{created at a
  particular site initially}.

To circumvent both problems, we add an auxiliary state space
$\mathcal{H}^a_i$ of the same dimension as $\mathcal{H}^p_i$ to each
site of our lattice. The total Hilbert space $\mathcal{H}$ is then
defined as the tensor product of the auxiliary and physical tensor
product spaces on each lattice site
\begin{equation}
  \mathcal{H} = \bigotimes_i \left( \mathcal{H}^p_i \otimes \mathcal{H}^a_i \right) \;.
\end{equation}
The initial quantum state $\Ket{\psi^p}$ is extended by a
suitably-chosen empty quantum state $\Ket{0^a}$ to form a state in the
full Hilbert space $\Ket{\psi} = \Ket{\psi^p} \otimes \Ket{0^a}$. In
the case of the $t-J$ model, for example, $\Ket{0^a}$ is the state
with zero particles on each site in the auxiliary system. The
Hamiltonian $\hat H$ used for the time evolution still only acts on
the physical system.

We then replace the excitation operator $\hat Y$ by a form which
conserves all symmetries of the system, namely
\begin{equation}
  \hat X = \prod_i \left( \hat 1 + \epsilon \hat x^p_i \left(\hat x^a_i\right)^\dagger + \mathrm{h.c.} \right) \;,
\end{equation}
where for convenience with existing implementations, we then instead
use the local exponential form
\begin{equation}
  \hat X = \prod_i \mathrm{exp} \left\{ \epsilon \hat x^p_i \left(\hat x^a_i\right)^\dagger + \mathrm{h.c.} \right\} \;.
\end{equation}
Instead of creating excitations from nothing as $\hat Y$ did, $\hat X$
now moves (e.g.) particles from the physical to the auxiliary system
and thereby creates an excitation in the physical sector. The density
of particles moved and hence the density of local excitations is given
by $\epsilon$, ideally we want to consider the case $\epsilon \to
0$.
No symmetry is broken during this process if we account for auxiliary
particles in the same way as we account for physical particles and
$\hat X$ hence leaves the fermionic parity of the state well-defined.

Additionally, it is now possible to not only post-select based on the
physical state of some particular site (to select an excitation
present there after the evolution), but also to post-select based on
the auxiliary state of some particular site. Because there are no
dynamics in the auxiliary layer, the auxiliary state at time $t$ is
equal to the auxiliary state at time $0$ and hence allows for the
selection of an excitation which was created at a particular site
initially.

\section{Application to the $t-J$ model}

Specifically, we consider the two-dimensional $t-J$ model on the
square lattice with a local physical three-dimensional state space
$\mathcal{H}^p_i = \mathrm{span}\left\{ \Ket{0^p_i},
  \Ket{\uparrow^p_i}, \Ket{\downarrow^p_i} \right\}.$
Taking a second such space $\mathcal{H}^a_i$ increases the local
physical dimension of the iPEPS tensor from three to nine, but iPEPS
methods scale favourably in this dimension, so this is not a
concern. Let $\hat c^{p(\dagger)}_{i\sigma}$ annihilate (create) a
physical fermion on site $i$ with spin $\sigma$, let
$\hat s^{p[+,-,z]}_i$ be the physical spin-$[+,-,z]$ operator on site
$i$ (0 if the site is empty) where $\hat s^z$ has eigenvalues
$\pm \nicefrac{1}{2}$ and let $\hat c^{a(\dagger)}_{i\sigma}$
annihilate (create) an auxiliary fermion on site $i$ with spin
$\sigma$. Finally, let $\hat n^p_i$ ($\hat n^a_i$) denote the particle
number operator ($0$ or $1$) on the physical (auxiliary) site $i$.

The Hamiltonian
\begin{equation}
  \hat H = -t\sum_{\langle i, j\rangle, \sigma} \left( \hat c^{p\dagger}_{i\sigma} c^p_{j\sigma} + \hat c^{p\dagger}_{j\sigma} c^p_{i\sigma} \right) + J \sum_{\langle i,j \rangle} \left[ \frac{1}{2} \left( \hat s^{p+}_i \hat s^{p-}_j + \hat s^{p+}_j \hat s^{p-}_i \right) + \hat s^{pz}_i \hat s^{pz}_j - \frac{1}{4} \hat n^p_i \hat n^p_j \right]
\end{equation}
acts on the physical sector only and is the standard $t-J$ Hamiltonian
linking all nearest-neighbour sites $\langle i, j \rangle$. Here, we
fix $t=1$ and $J=\nicefrac{1}{3}$.

Now take $\ket{\mathrm{GS}}$ to be an approximation of the infinite ground
state of $\hat H$ at a given iPEPS bond dimension $D$ and half-filling
(one fermion per site) in the physical sector, with the auxiliary
sector being entirely empty:
\begin{align}
  \ket{\mathrm{GS}} = \ket{\mathrm{GS}^p} \otimes \ket{0^a} \;.
\end{align}
The physical ground state $\ket{\mathrm{GS}^p}$ is simply the
ground-state of the Heisenberg Hamiltonian, which can be reasonably
well approximated by a $D=4$ or $D=5$ iPEPS (other states may of
course require a larger bond dimension). This state breaks
translational invariance, so we use a $2 \times 2$ unit cell. It
preserves both $\mathrm{U}(1)_N$ particle number and
$\mathrm{U}(1)_{S^z}$ spin-projection symmetry and we make use of
both\cite{hubig18:_abelian}. Fermionic commutation relations are
ensured using the fermionic tensor network
ansatz\cite{barthel09:_contr, bultinck17:_fermion} as implemented in
\textsc{SyTen}'s \texttt{STensor}
class\cite{hubig17:_symmet_protec_tensor_networ, hubig:_syten_toolk}.

Given $\Ket{\mathrm{GS}}$ as described above, we create the initial
excitation with the operator
\begin{equation}
  \hat X = \prod_i \mathrm{exp}\left\{ \epsilon \sum_\sigma \left( \hat c^{p\dagger}_{i\sigma} c^a_{i\sigma} + \hat c^{a\dagger}_{i\sigma} c^p_{i\sigma} \right) \right\} \;.
\end{equation}
This operator will move particles from the occupied physical sector to
the empty auxiliary sector and results in new state $\Ket{\psi(0)}$
with a finite hole density on each physical site. Evolving this state
under the physical Hamiltonian $\hat H$ is straightforward and for a
given time $t$ results in a state
\begin{equation}
  \Ket{\psi(t)} = e^{-\mathrm{i}t\hat H} \Ket{\psi(0)}\;.
\end{equation}
In the following, we are particularly interested in (a) the return
probability $p^R(t)$ of a hole to its creation site and (b) the
diagonal spin-spin correlator $z^{\mathrm{diag}}(t)$ at time $t$ with
a hole present at time $t$ between the two spins.

The return probability $p^R(t)$ is given by
\begin{equation}
  p^R(t) = \frac{\Bra{\psi(t)} \left( \hat 1 - \hat n^p_i \right) \hat n^a_i \Ket{\psi(t)}}{\Braket{\psi(t) | \hat n^a_i | \psi(t)}} \;,
\end{equation}
where the numerator evaluates the joint probability of a hole created
at site $i$ (via the density on the auxiliary site, $\hat n^a_i$)
present there at a later time (via the density on the physical site,
$\hat n^p_i$) with the denumerator conditioning on the initial
creation of a hole at this site. As the hole density is low, we
neglect the case of the hole created at site $i$ moving away and
another hole created at some neighbouring site $j$ taking its place.

For the diagonal spin-spin correlator around a hole, let us first
define site indices $00$, $10$ and $11$ of the $2 \times 2$ unit
cell. The correlator is then
\begin{align}
  z^\mathrm{diag}(t) & = \frac{\Bra{\psi} \hat s^{pz}_{00}(t) \left( \hat 1 - \hat n^p_{10}(t) \right) \hat s^{pz}_{11}(t)\Ket{\psi}}{\Bra{\psi} \left( \hat 1 - \hat n^p_{10}(t) \right) \Ket{\psi}} \\
                         & = \frac{\Bra{\psi(t)} \hat s^{pz}_{00} \left( \hat 1 - \hat n^p_{10} \right) \hat s^{pz}_{11}\Ket{\psi(t)}}{\Bra{\psi(t)} \left( \hat 1 - \hat n^p_{10} \right) \Ket{\psi(t)}} \;.
\end{align}

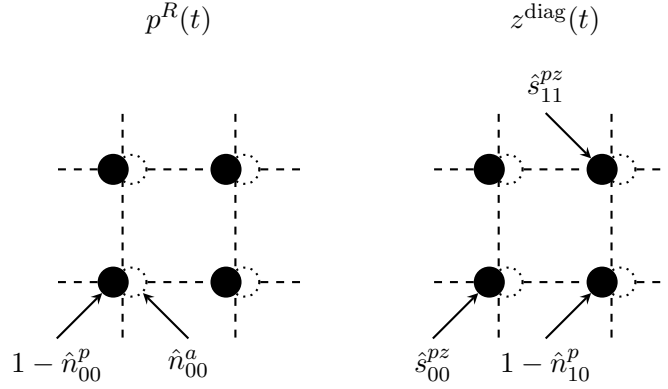
\begin{figure}
  \centering
  \tikzsetnextfilename{observables-tn}
  \begin{tikzpicture}
    \node[] (pr) at (0.875,3.5) {$p^R(t)$};

    \node[draw,circle,dotted,draw=black, fill=white, thick, minimum width=1em] (aux00) at (0.25,0) {};
    \node[draw,circle,draw=black, fill=black,thick, minimum width=1em] (phys00) at (0,0) {};

    \node[draw,circle,dotted,draw=black, fill=white, thick, minimum width=1em] (aux01) at (0.25,1.5) {};
    \node[draw,circle,draw=black, fill=black,thick, minimum width=1em] (phys01) at (0,1.5) {};

    \node[draw,circle,dotted,draw=black, fill=white, thick, minimum width=1em] (aux10) at (1.75,0) {};
    \node[draw,circle,draw=black, fill=black,thick, minimum width=1em] (phys10) at (1.5,0) {};

    \node[draw,circle,dotted,draw=black, fill=white, thick, minimum width=1em] (aux11) at (1.75,1.5) {};
    \node[draw,circle,draw=black, fill=black,thick, minimum width=1em] (phys11) at (1.5,1.5) {};

    \draw[dashed,thick] (0.125,0) -- (0.125,1.5);
    \draw[dashed,thick] (1.625,0) -- (1.625,1.5);
    \draw[dashed,thick] (aux00) -- (phys10);
    \draw[dashed,thick] (aux01) -- (phys11);
    \draw[dashed,thick] (phys00) -- +(-0.75,0);
    \draw[dashed,thick] (phys01) -- +(-0.75,0);
    \draw[dashed,thick] (aux10) -- +(+0.75,0);
    \draw[dashed,thick] (aux11) -- +(+0.75,0);
    \draw[dashed,thick] (0.125,0) -- +(0,-0.75);
    \draw[dashed,thick] (0.125,1.5) -- +(0,+0.75);
    \draw[dashed,thick] (1.625,1.5) -- +(0,+0.75);
    \draw[dashed,thick] (1.625,0) -- +(0,-0.75);

    \draw[thick,<-] (phys00) -- +(-0.75,-0.75) node[below]{$1 - \hat n^p_{00}$};
    \draw[thick,<-] (aux00) --  +(0.75,-0.75) node[below]{$\hat n^a_{00}$};

    \node[] (zt) at (5.875,3.5) {$z^{\mathrm{diag}}(t)$};

    \node[draw,circle,dotted,draw=black, fill=white, thick, minimum width=1em] (aux00) at (5.25,0) {};
    \node[draw,circle,draw=black, fill=black,thick, minimum width=1em] (phys00) at (5,0) {};

    \node[draw,circle,dotted,draw=black, fill=white, thick, minimum width=1em] (aux01) at (5.25,1.5) {};
    \node[draw,circle,draw=black, fill=black,thick, minimum width=1em] (phys01) at (5,1.5) {};

    \node[draw,circle,dotted,draw=black, fill=white, thick, minimum width=1em] (aux10) at (6.75,0) {};
    \node[draw,circle,draw=black, fill=black,thick, minimum width=1em] (phys10) at (6.5,0) {};

    \node[draw,circle,dotted,draw=black, fill=white, thick, minimum width=1em] (aux11) at (6.75,1.5) {};
    \node[draw,circle,draw=black, fill=black,thick, minimum width=1em] (phys11) at (6.5,1.5) {};

    \draw[dashed,thick] (5.125,0) -- (5.125,1.5);
    \draw[dashed,thick] (6.625,0) -- (6.625,1.5);
    \draw[dashed,thick] (aux00) -- (phys10);
    \draw[dashed,thick] (aux01) -- (phys11);
    \draw[dashed,thick] (phys00) -- +(-0.75,0);
    \draw[dashed,thick] (phys01) -- +(-0.75,0);
    \draw[dashed,thick] (aux10) -- +(+0.75,0);
    \draw[dashed,thick] (aux11) -- +(+0.75,0);
    \draw[dashed,thick] (5.125,0) -- +(0,-0.75);
    \draw[dashed,thick] (5.125,1.5) -- +(0,+0.75);
    \draw[dashed,thick] (6.625,1.5) -- +(0,+0.75);
    \draw[dashed,thick] (6.625,0) -- +(0,-0.75);

    \draw[thick,<-] (phys00) -- +(-0.75,-0.75) node[below]{$\hat s^{pz}_{00}$};
    \draw[thick,<-] (phys11) -- +(-0.75,+0.75) node[above]{$\hat s^{pz}_{11}$};
    \draw[thick,<-] (phys10)  -- +(-0.75,-0.75) node[below]{$1-\hat n^p_{10}$};

  \end{tikzpicture}
  \caption{\label{fig:observables-tn}Top view of a single iPEPS unit
    cell, representing a state $\Ket{\psi(t)}$. Each site is the
    product space of a physical (black) and auxiliary (white/dotted)
    site. Sites are connected via iPEPS virtual bonds (dashed). Left:
    The return probability $p^R(t)$ is evaluated by measuring
    $1 - \hat n^p_{i}$ and $\hat n^a_i$ at the same iPEPS site. Right:
    The equal-time correlator $z^\mathrm{diag}(t)$ around a hole
    at time $t$ is evaluated by measuring $\hat s^{pz}_{00}$,
    $\hat s^{pz}_{11}$ and $1 - \hat n^p_{10}$.}
\end{figure}

These correlators are sketched in \Cref{fig:observables-tn}. Note
that, if desired and with larger computational effort, it would be
conceivable to repeat the same calculation at different values of
$\epsilon$ and subsequently extrapolate $\epsilon \to 0$.

\section{Results}

In the following, we apply the method described above to evaluate the
return probability and diagonal-nearest-neighbour spin correlators in
the $t-J$ model after the effective introduction of a single hole. We
also simulate this system using time-dependent matrix-product
states\cite{paeckel19:_time} on cylinders of width 4 and 6 to obtain
comparison data for short times.

\begin{figure}
  \centering
  \includegraphics[width=0.8\textwidth]{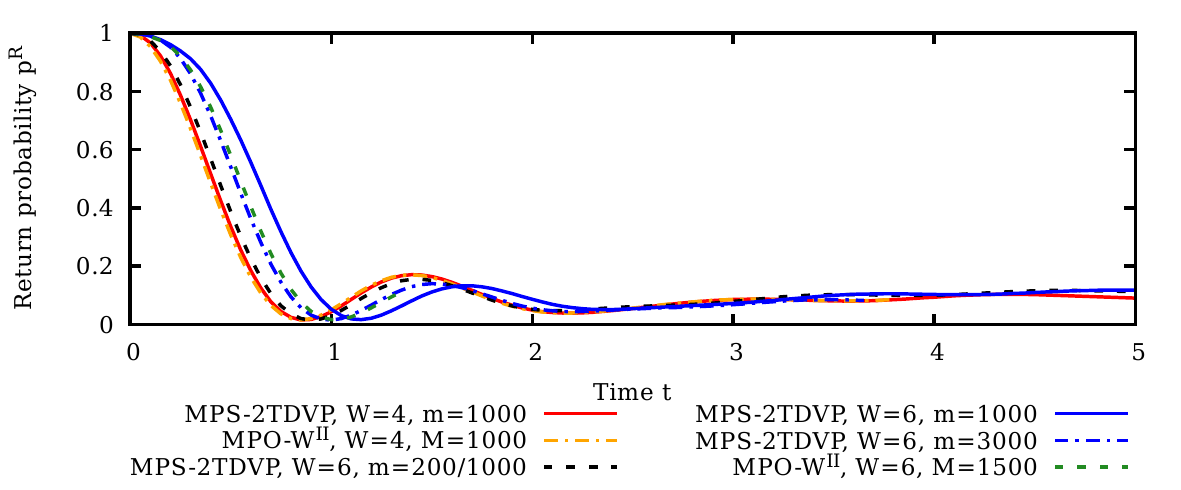}
  \caption{\label{fig:retprob-tdmps}Return probability as calculated
    using MPS-TDVP or the MPO $W^{\mathrm{II}}$ methods at
    $J=\nicefrac{1}{3}$. Both methods used a step size
    $\delta t = 0.05$. On $W=4$ cylinders, results are well-converged
    at $m=1000$ already. On $W=6$ cylinders, we only achieve
    qualitative convergence as the required MPS bond dimension would
    exceed computational resources.}
\end{figure}

\paragraph{Time-dependent matrix-product states} on cylindrical
geometries are used to provide comparison data, assumed to be valid at
least for short times when the finite circumference of the cylinders
is not yet relevant. We compute the ground-states of the $t-J$ model
at half-filling and apply an excitation
$\hat c_{0,\uparrow} + \hat c_{0,\downarrow}$ in the centre of the
system. The resulting excited state is then time-evolved with either
the 2TDVP\cite{haegeman16:_unify} or the MPO $W^{\mathrm{II}}$
method\cite{kjaell13:_phase_xxz, zaletel15:_time,
  gohlke17:_dynam_kitaev_heisen_model} using the
\textsc{SyTen}\cite{hubig17:_symmet_protec_tensor_networ,
  hubig:_syten_toolk} and \textsc{TeNPy}
toolkits\cite{hauschild18:_effic_tensor_networ} respectively. The
return probability is given simply as $\Braket{1 - \hat n_0(t)}$. On
cylinders of width $W=4$, convergence is easy to achieve at modest
bond dimensions $m=1000$, increasing the bond dimension further (up to
$m=5000$) does not lead to different results. As the MPS bond
dimension scales exponentially with the circumference of the cylinder,
convergence is more difficult on $W=6$ cylinders. Running the time
evolution at the same fixed bond dimension as the initial ground state
does not converge well. Preparing the initial ground state at a
smaller bond dimension $200$ and then running the time evolution at
bond dimension $m=1000$ leads to results at least on short times very
similar to the $W=4$ cylinder (cf.~\cref{fig:retprob-tdmps}), which is
expected as the short-time dynamics are independent of the spin
background and hence governed by the hole motion only. Departing from
the short-time regime, however, the results become
uncontrolled. Increasing the bond dimension further or evolving with
the same bond dimension as the initial state does not lead to good
convergence. Additionally, while the hole spreads isotropically along
the $x$- and $y$-direction on the $W=4$ cylinder, this is not the case
on the $W=6$ cylinder (not shown). Overall, we only obtain reliable
data for the return probability on cylinders of width $W=4$ and qualitative data for cylinders of width $W=6$.

\begin{figure}[p]
  \centering
  \includegraphics[width=0.8\textwidth]{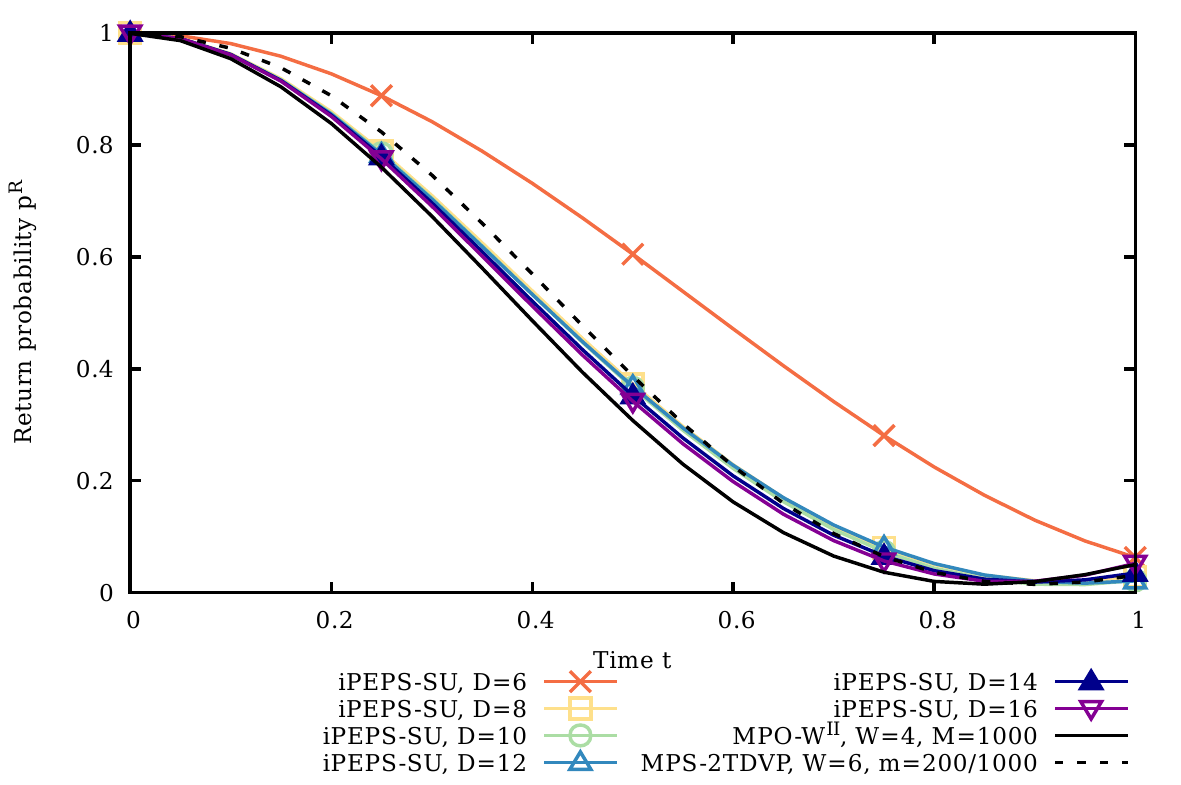}
  \caption{\label{fig:retprob_short}Return probability $p^R(t)$
    calculated using iPEPS with the simple update and td-MPS on short
    times from an initial $D^\prime=4$ state excited with a global
    hole density of 0.01 and $J=\nicefrac{1}{3}$ with various iPEPS
    bond dimensions $D$. We observe good convergence of the initial
    decay once $D \geq 8$. Data is evaluated every $\delta t = 0.05$,
    with symbols shown only for identification.}
\end{figure}

\begin{figure}[p]
  \centering
  \includegraphics[width=0.8\textwidth]{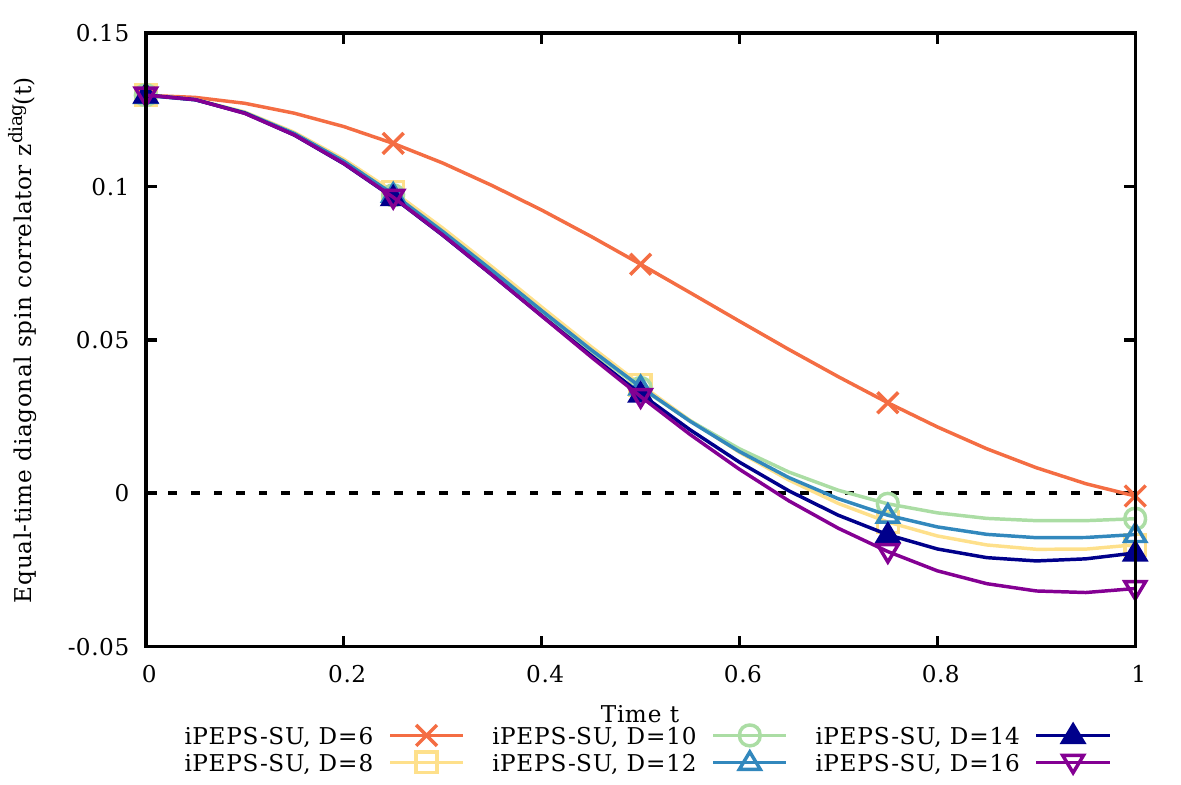}
  \caption{\label{fig:ztptzt_short}Equal-time diagonal spin correlator
    $z^\mathrm{diag}(t)$ when a hole is present in the lower right
    side of the two spins calculated using iPEPS with the simple
    update. The expected zero crossing is observed when increasing the
    iPEPS bond dimension around time $t \approx 0.6$. Data is
    evaluated every $\delta t = 0.05$, with symbols shown only for
    identification.}
\end{figure}

\paragraph{In the iPEPS simulation,} we use the fast full update (FFU,
\cite{phien15:_fast, phien15:_infin}) to obtain the initial ground
state and perform the subsequent evolution with the simple update
(SU). While the (fast) full update would be able to make better use of
the bond dimension of our state, we have encountered some stability
issues\cite{hubig19:_time} resulting from this update method which
lead to very limited time scales. The simple update may not make
perfect use of the iPEPS bond dimension but, given a sufficiently
large bond dimension, still provides good results without any of the
stability issues observed with the FFU.

We prepare the initial (ground) state at an initial bond dimension
$D^\prime = 4$ and create an excitation density of $10^{-2}$. During
the subsequent real-time evolution, we allow a range of bond
dimensions $D=4, \ldots, 16$. We focus on even bond dimensions $D$, as
odd bond dimensions show slightly worse convergence behaviour due to
truncation within spin multiplets. Future computational and
algorithmic advances may make bond dimensions $D > 17$ possible.  We
use a time step size $\delta t = 0.01$ together with a second-order
Trotter decomposition of the time-evolution operator.

Exploratory calculations at $D^\prime = 5$ and/or hole density
$\approx 10^{-4}$ result in decreased hole mobility at a given
evolution bond dimension $D$ as the competition between spin and hole
entanglement during the iPEPS state truncation favour the spin sector
disproportionally when it is initially more strongly entanglend
($D^\prime = 5$) or there are fewer holes. Hole mobility still
increases when increasing the evolution bond dimension $D$, but
convergence is much slower than when starting with $D^\prime = 4$.

Expectation values are calculated using the corner transfer matrix at
increasing bond dimensions $\chi$ until the difference between results
of two successive dimensions $\chi$ and $2 \chi$ are sufficiently
small; error bars are smaller than symbol sizes in all cases.

\Cref{fig:retprob_short} and \Cref{fig:ztptzt_short} show the
short-time dynamics of the return probability $p^R(t)$ and diagonal
spin-spin correlator $z^\mathrm{diag}(t)$ calculated with iPEPS. We
observe good convergence in the bond dimension starting from
$D \geq 8$ for short times. There, the td-MPS results are
reproduced. In particular, the motion of the hole away from its
initial site on times of the order of the nearest-neighbour hopping is
captured well. At the same time, $z^\mathrm{diag}(t)$ becomes negative
because the moving hole distorts the original antiferromagnetic
background. Hence, spin correlators between both originally
nearest-neighbour and originally next-nearest-neighbour fermions
contribute to $z^\mathrm{diag}(t)$. The stronger nearest-neighbour
correlators then dominate the sum and cause the observed sign
change. Because the $\mathrm{SU}(2)$-spin symmetry is spontaneously
broken along the preferred $z$-axis in the iPEPS calculation but still
present in the finite td-MPS calculations, a comparison of numerical
values is not meaningful in this case.

\begin{figure}
  \centering
  \includegraphics[width=0.8\textwidth]{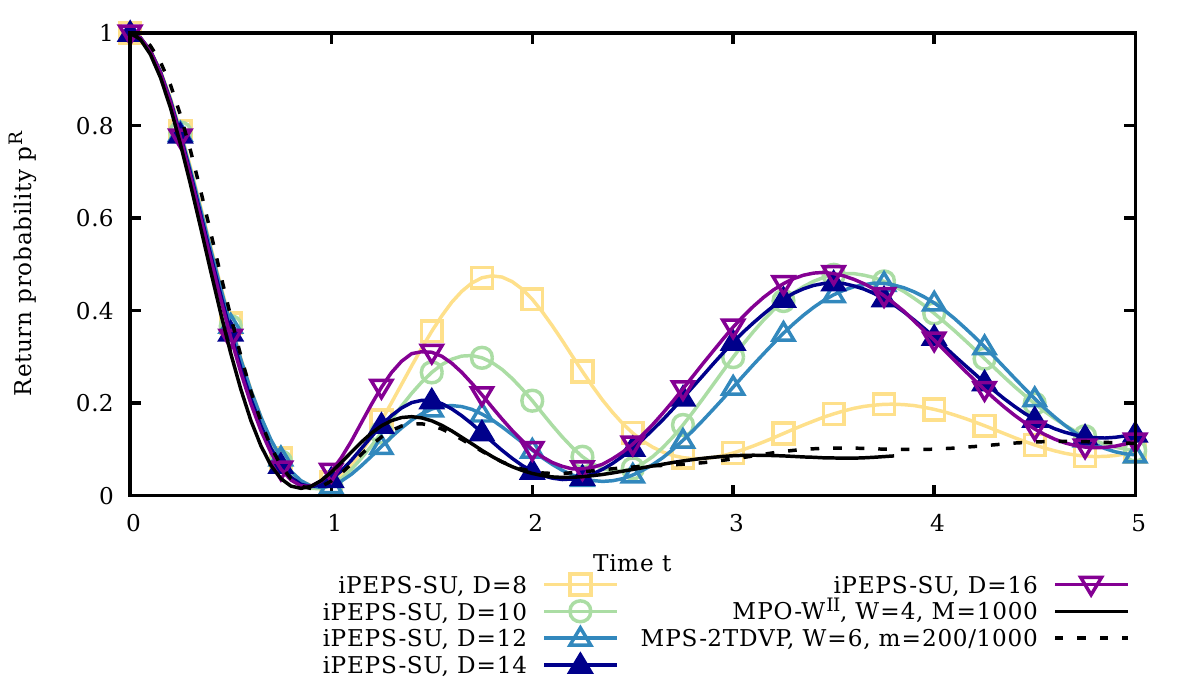}
  \caption{\label{fig:retprob}Same as \Cref{fig:retprob_short} for
    longer times $t \geq 1$. The return probability shows qualitative
    features common to all calculations at large bond dimensions, but
    quantitative convergence is difficult. The revival around
    $t \approx 3.5$ is not expected and likely due to limited
    entanglement in our ansatz.}
\end{figure}

For longer times, convergence is very difficult, as our ansatz is
inherently limited in entanglement and -- due to the simple update --
does not make optimal use of the available bond dimension.\footnote{A
  further check on convergence may lie in a deeper analysis of the
  singular value spectrum obtained after each simple update. While not
  exact due to missing normalisation of the environment, one might
  still expect a flattening of the spectrum as entanglement grows over
  time. We would like to thank Referee 3 for this suggestion.}
However, the first revival of the return probability observed in the
td-MPS data is still reproduced well by the iPEPS results around
$t \approx 1.5$, cf.~\Cref{fig:retprob}. The iPEPS data also contains
a second, much larger revival at later times $t \approx 3.5$ which is
not observed in the td-MPS data and not physically expected either
(instead we expect the hole to move away from its creation point with
frustrated spins left behind healed by spin
flips\cite{bohrdt19:_dynam}). At the moment, it is unclear whether
this revival is due to limited entanglement in the iPEPS ansatz which
hinders healing of frustrated spins through spin-exchange interactions
and hence increases the cost of moving the hole further from its
origin or a side-effect of the typically overestimated magnetisation
in the iPEPS ground state which may lead to more Ising-like physics.

\section{Conclusion}

We have shown that both the simulation of local excitations and the
evaluation of time-dependent correlators is possible within the iPEPS
formalism.  Our predictions, such as the sign-change of diagonal
correlators around the hole in \Cref{fig:ztptzt_short}, can already be
tested in state-of-the-art quantum-gas
microscopes\cite{mazurenko17:_fermi, chiu19:_strin_hubbar,
  koepsell19:_imagin_fermi,
  vijayan19:_time_resol_obser_spin_charg}. Future work using an
environment-based truncation scheme such as the FFU together with a
stabilised environment (e.g. as introduced in
Ref.~\cite{vanderstraeten16:_gradien}) will be in a position to make
much better use of the available bond dimension than the simple update
employed here and hence will be able to analyse the physics of the
system for longer times, in particular the interactions between holons
and spinons. This would also open an alternative
avenue\cite{vanderstraeten19:_simul} to obtaining spectral functions
of two-dimensional systems.

\section*{Acknowledgements}

The authors would like to thank I. Bloch, E. Demler, D. Golez,
M. Greiner, I. P. McCulloch, F. Pollmann, and U. Schollwöck for useful
discussions.

\paragraph{Funding information}
C. H. and J. I. C. acknowledge funding through ERC Grant QUENOCOBA,
ERC-2016-ADG (Grant no. 742102) by the DFG under Germany's Excellence
Strategy -- EXC-2111 -- 390814868. A.B., F.G., and M.K. acknowledge
support from the Technical University of Munich -- Institute for
Advanced Study, funded by the German Excellence Initiative, the
European Union FP7 under grant agreement 291763, the Deutsche
Forschungsgemeinschaft (DFG, German Research Foundation) under
Germany's Excellence Strategy -- EXC-2111 -- 390814868, DFG grant
No. KN1254/1-1, DFG TRR80 (Project F8), and from the European Research
Council (ERC) under the European Union's Horizon 2020 research and
innovation programme (grant agreement No. 851161).

\end{document}